\documentclass[preprint]{aastex62}
\usepackage{amsmath,amstext}
\usepackage{graphicx}

\newcounter{RomanNumber}

\received{****}
\revised{****}
\accepted{****}

\shorttitle{Emissions of Leaky Fast Sausage Modes}
\shortauthors{Shi et al.}
\begin{document}

\title{Synthetic Extreme-ultraviolet Emissions Modulated by Leaky Fast Sausage Modes in Solar Active Region Loops}

\correspondingauthor{Mijie Shi}
\email{shimijie@sdu.edu.cn}

\author{Mijie Shi}
\affiliation{Shandong Provincial Key Laboratory of Optical Astronomy and Solar-Terrestrial Environment, Institute of Space Sciences, Shandong University, Weihai 264209, China}
\affiliation{CAS Key Laboratory of Solar Activity, National Astronomical Observatories, Beijing 100012, China}

\author{Bo Li}
\affiliation{Shandong Provincial Key Laboratory of Optical Astronomy and Solar-Terrestrial Environment, Institute of Space Sciences, Shandong University, Weihai 264209, China}

\author{Zhenghua Huang}
\affiliation{Shandong Provincial Key Laboratory of Optical Astronomy and Solar-Terrestrial Environment, Institute of Space Sciences, Shandong University, Weihai 264209, China}

\author{Shao-Xia Chen}
\affiliation{Shandong Provincial Key Laboratory of Optical Astronomy and Solar-Terrestrial Environment, Institute of Space Sciences, Shandong University, Weihai 264209, China}



\begin{abstract}
We study the extreme-ultraviolet (EUV) emissions modulated by leaky fast sausage modes (FSMs)
	in solar active region loops and examine their observational signatures via spectrometers like EIS. 
After computing fluid variables of leaky FSMs with MHD simulations,
	we forward-model the intensity and spectral properties of the Fe X 185~\AA~and Fe XII 195~\AA~lines
	by incorporating non-equilibrium ionization (NEI) in the computations of the relevant ionic fractions.
The damping times derived from the intensity variations are then compared with the wave values,
namely the damping times directly found from our MHD simulations.
Our results show that
in the equilibrium ionization cases, the density variations and the intensity variations
	can be either in phase or in anti-phase, depending on the loop temperature. 
NEI considerably impacts the intensity variations
	but has only marginal effects on the derived Doppler velocity or Doppler width.
We find that the damping time derived from the intensity can largely reflect the wave damping time
	if the loop temperature is not drastically different from the nominal formation temperature of the corresponding emission line.
These results are helpful for understanding the modulations to the EUV emissions by leaky FSMs and 
	hence helpful for identifying FSMs in solar active region loops.

\end{abstract}




\section{Introduction} 
Magnetohydrodynamic (MHD) waves and oscillations abound in the structured
	solar atmosphere 
	\cite[see reviews, e.g.,][]{2005LRSP....2....3N,2007SoPh..246....3B,2012RSPTA.370.3193D}.
Fast sausage modes (FSMs) are a kind of MHD waves characterized by the axial symmetry of 
	the associated perturbations 
	\citep{1983SoPh...88..179E}.
Depending on the longitudinal wavenumber,
	FSMs can be either leaky or trapped
	\citep{1986SoPh..103..277C}.
FSMs can periodically compress the loop and are thus thought of as one of the mechanisms 
	accounting for quasi-periodic pulsations (QPPs) in solar flares
	\cite[e.g.,][]{2009SSRv..149..119N,2016SoPh..291.3143V}. 
In terms of observations, signatures of FSMs had been detected in flare loops in the radio band
	by Nobeyama Radioheliograph \citep[e.g.,][]{2005A&A...439..727M,2015A&A...574A..53K},
	in EUV by SDO/AIA \citep{2012ApJ...755..113S} and IRIS \citep{2016ApJ...823L..16T},
	as well as in X-ray emissions by RHESSI \citep{2010SoPh..263..163Z}.
FSMs were also shown to be associated with some fine structures of radio bursts
	observed by the Chinese Solar Broadband Radio Spectrometer \citep{2013ApJ...777..159Y},
	and the Assembly of Metric-band Aperture Telescope and Real-time Analysis System \citep{2018ApJ...855L..29K}.

However, confidently identifying FSMs is not often straightforward.
For this reason, a number of forward modeling studies have been carried out in the literature 
	in order to obtain their observational signatures.
\cite{2003A&A...409..325C} and \cite{2012A&A...543A..12G} integrated the squared density along a line of sight 
	to examine the geometrical effects on the observability of FSMs.
\cite{2013A&A...555A..74A} took into account the contribution function 
	for the computation of emissivity, 
	and examine the geometrical and instrumental effects on the EUV emissions modulated by FSMs.
Also for the EUV emissions, our recent works analyzed the intensity and spectral properties of FSMs in both 
	active region (AR) loops \citep{2019ApJ...870...99S}
	and flare loops \citep{2019ApJ...874...87S}
	by taking into account the non-equilibrium ionization (NEI) effects.
 It turns out that NEI needs to be considered given that the periods of FSMs tend not to be much longer than the relevant ionization and/or recombination timescales.
In the radio band, \cite{2014ApJ...785...86R} analyzed the gyrosynchrotron intensity modulated by FSMs.
Most of these forward modeling works focused on the emissions of FSMs in the trapped regime.
During these analyses, some properties of the observed quantities, 
	such as the periods, phase relations, and modulation amplitudes
	were extensively examined.
However, for typical AR loop parameters, FSMs are usually in the leaky regime,
	experiencing the apparent wave attenuation as a result of lateral leakage 
	\citep[see e.g.,][]{1982SoPh...75....3S, 1986SoPh..103..277C}.
Forward modeling analysis of damped waves has been conducted by
	\cite{2008SoPh..252..101D}, but for propagating slow magnetoacoustic waves.
They found that the damping rates derived from the intensities do not often reflect the wave values,
	and in some cases, even the periods can be different.

This study aims to examine the emission properties of leaky FSMs in the EUV band
	by considering NEI when forward modeling the MHD simulated data.
To our knowledge, this kind of forward modeling analysis for leaky FSMs has not been conducted,
and the outcomes are important to identify the leaky FSMs in AR loops.
We first perform MHD simulations of leaky FSMs,
	and then forward model the emission properties of the Fe X 185 \AA~and Fe XII 195 \AA~lines,
	taking into account both the spatial and spectral resolutions of the EIS instrument onboard Hinode.
The numerically simulated leaky FSMs are shown in Section \ref{sec_model_formulation}.
We present the forward modeling results in Section \ref{sec_forward_model}
	and conclude this work in Section \ref{sec:summary_and_conclusion}.

\section{Numerically simulated Fast Sausage Modes}
\label{sec_model_formulation}
We model an AR loop as an axisymmetric straight cylinder.
In a standard cylindrical coordinate system,
	both the cylinder and the equilibrium magnetic field are directed in the $z$-direction.
The equilibrium parameters are assumed to be a function of $r$ only.
To be specific, the transverse distribution of the electron number density $N$ reads
\begin{eqnarray}
\label{equ:den}
N(r,t=0) = \frac{N_i+N_e}{2}-\frac{N_i-N_e}{2}{\rm tanh}(\frac{r-R_0}{\delta}),
\end{eqnarray}
where $N_i=10^9~{\rm cm^{-3}}~(N_e=10^8~\rm cm^{-3})$ denotes the electron density in the interior (exterior) of the loop.
Furthermore, $R_0 = 2~\rm{Mm}$ represents the mean radius of the loop,
	and $\delta=R_0/4$ is a parameter controlling the steepness of the density distribution.
The electron temperatures inside and outside of the loop are denoted by $T_i$ and $T_e$, respectively.
In all simulations, $T_e$ is fixed at 0.7 MK,
while $T_i$ is allowed to vary between 0.9 MK and 1.7 MK. 
The transverse profile of the thermal pressure follows the same $r$-dependence as the density,
	with $N_{i(e)}$ in Equation \eqref{equ:den} replaced by $p_{i(e)}$, 
where $p_{i(e)}=2N_{i(e)}k_BT_{i(e)}$ is the thermal pressure inside (outside) of the loop.
The transverse distribution of the magnetic field is then derived via the transverse force balance condition. 
The interior and exterior magnetic fields are $[B_i,B_e] = [10.87,11.38]$ G,
and the Alfv\'en speeds are $[v_{Ai},v_{Ae}] = [750,2483]~{\rm km~s^{-1}}$.
The plasma $\beta$ at the loop axis ($\beta_i$) varies between $0.026$ and $0.05$ for different choices of $T_i$,
while $\beta_e = 0.002$ is a fixed value outside the loop.
The loop length is $L_0 = 50 R_0$.

	Both the physical and geometrical parameters for the equilibrium are compatible with typical active region loops 
	~\citep[e.g.,][]{2004ApJ...600..458A, 2007ApJ...662L.119S}. 
	However, by far the majority of the sausage modes reported in the literature are associated with flaring loops, 
	which possess a higher density and a larger minor radius ~\citep[e.g.,][]{2018SSRv..214...45M}. 
	In active region loops, convincing observational evidence has yet to be found for the existence of sausage modes.
	One reason for this lack of evidence is that standing sausage waves, at least the fundamental modes, 
	are likely to experience rather significant attenuation due to lateral leakage~\citep[e.g.,][and references therein]{2018JPhA...51b5501C}.
	Therefore one primary objective of this manuscript is to examine the observational signatures of leaky modes, 
	thereby offering some guidance for identifying sausage modes in active region loops with future spectrometers similar to Hinode/EIS but with some better temporal cadence.

	Before proceeding, it is interesting to note that for the wave properties themselves, 
	a finite plasma beta plays an at most marginal role in determining the periods and damping times as long as these are measured in units of the transverse fast time rather than the transverse Alfv\'en time 
	(\citeauthor{2016ApJ...833..114C}~\citeyear{2016ApJ...833..114C}, 
	\citeauthor{2018ApJ...855...47C}~\citeyear{2018ApJ...855...47C};
	see also \citeauthor{2009A&A...503..569I}~\citeyear{2009A&A...503..569I}). 
	However, a finite temperature (and hence a finite plasma beta) is essential for synthesizing the (E)UV emissions, 
	primarily because of the sensitive temperature dependence of the contribution function. 
	A number of values for the electron temperature inside the loop are therefore examined to illustrate this effect.

To trigger FSMs, we set the initial transverse velocity perturbation as
\begin{eqnarray}
\label{equ:veloc_perturb}
v_r(r,t=0) = a_0v_{Ai}\sqrt{2}\frac{r}{\Delta}{\rm exp}(-\frac{r^2}{\Delta^2}+0.5){\rm sin}(k_0z),
\end{eqnarray}
where $a_0 = 0.04$ is the dimensionless magnitude of the velocity perturbation,
$\Delta$ dictates the width of the velocity perturbation and is fixed at $1.2R_0$,
and $k_0=\pi/L_0$ is the longitudinal wavenumber of the fundamental standing mode.
Equation \eqref{equ:veloc_perturb} ensures that the maximum velocity perturbation is $0.04v_{Ai}$. 

Appropriate boundary conditions (BCs) are necessary for the generation of standing FSMs.
For this reason, we use the same boundary conditions as in \cite{2016ApJ...833..114C}.
To be specific, at $r=0$, the BC reads
\begin{eqnarray}
v_r =B_r=0,~
\frac{\partial\rho}{\partial r}=
\frac{\partial v_z}{\partial r}=
\frac{\partial B_z}{\partial r}=
\frac{\partial p}{\partial r}=0.
\end{eqnarray}
The BC at $r=r_M$ is specified as "outflow".
We set $r_M$ to be sufficiently large 
so that in the simulated time interval the perturbations 
reflected off this boundary do not contaminate the inner region.
At the boundaries of $z=0$ and $L_0$,
all physical quantities are fixed at their initial values, except for $v_z$ and $B_r$
\begin{eqnarray}
\frac{\partial v_z}{\partial z}=
\frac{\partial B_r}{\partial z}=0.
\end{eqnarray}

\begin{figure}
  \begin{centering}
  \includegraphics[width=0.6\linewidth]{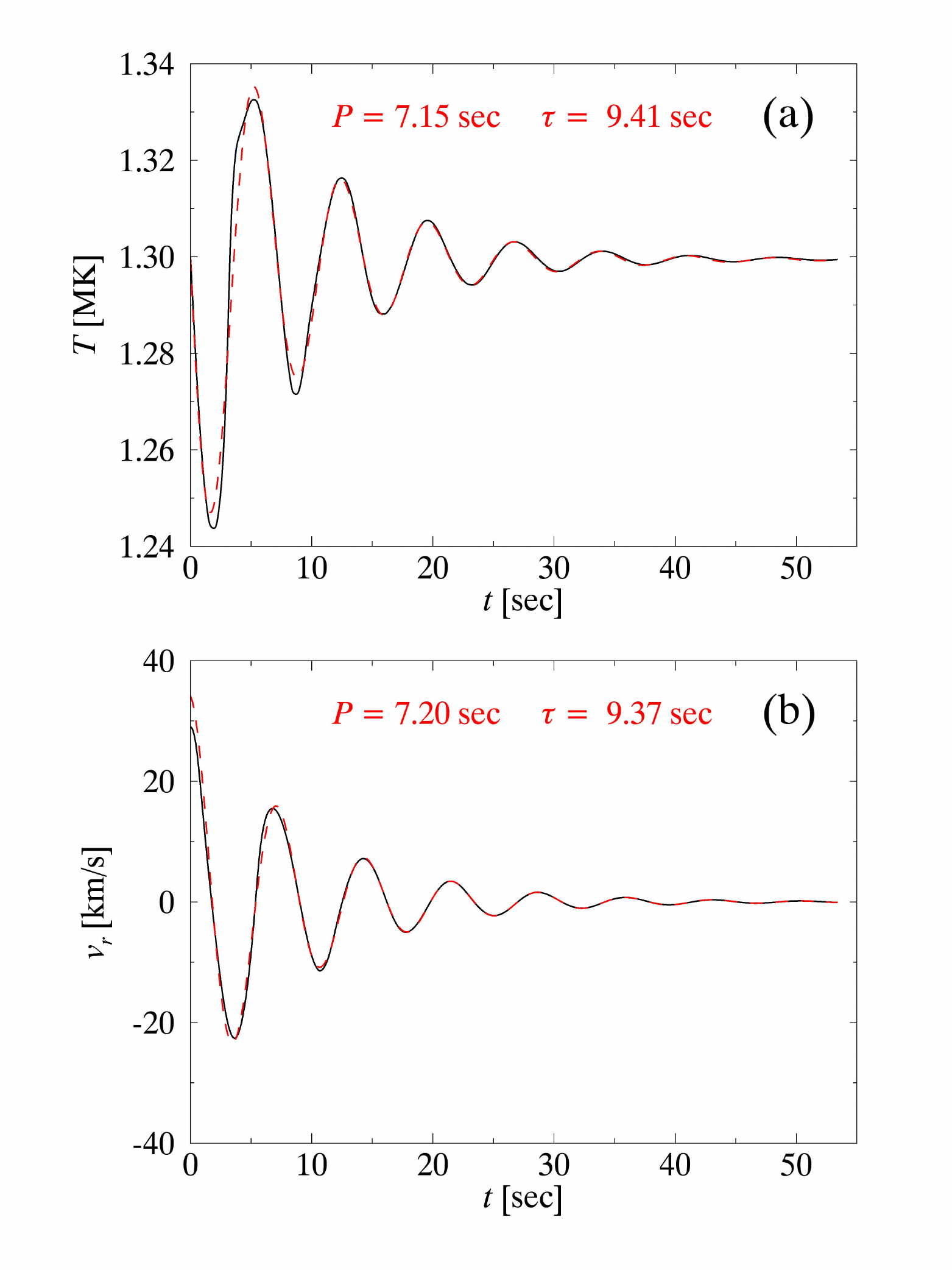}
  \caption{Temporal evolutions of (a) the temperature $T$ at $[r,z]=[0,L_0/2]$, and
  			(b) the transverse velocity $v_r$ at $[r,z]=[R_0,L_0/2]$ of leaky FSMs.
  			The red dashed lines show the fitting curves from Equation \eqref{equ:sine-exponential}, 
  			with the derived $P$ and $\tau$ labeled in each panel. 
  			Here the base model is examined, i.e., the electron temperature at the loop axis $T_i=1.3$~MK.}
  \label{f1}
 \end{centering}
\end{figure}

We perform our simulations using the ideal MHD module of the 
	PLUTO code \citep{2007ApJS..170..228M}.
Figure \ref{f1} shows the temporal evolutions of two fluid parameters for the simulation case 
	with $T_i = 1.3$~MK (referred to as the \textit{base model} hereafter).
To demonstrate the leaky property of the FSM,
we show in Figure \ref{f1}a the temperature evolution at the loop apex $([r,z]=[0,L_0/2])$
	and in Figure \ref{f1}b the transverse velocity evolution at $[r,z]=[R_0,L_0/2]$ by the black solid lines,
together with their fitting plots in the red dashed lines via equation
\begin{eqnarray}
\label{equ:sine-exponential}
F(t) = A_0 + A_1{\rm{sin}}(2\pi t/P+\phi){\rm{exp}}(-t/\tau),
\end{eqnarray}
where $A_0$ and $A_1$ are fitting constants, $\phi$ the phase shift,  
$P$ the period, and $\tau$ the damping time.

The fluid parameters of the base model and other cases (not shown) all show periodic oscillations with their amplitudes exponentially decreasing. 
These are the leaky FSMs.
In the next section, based on the simulated data of the leaky FSMs, 
 we forward model the emissions of the Fe X 185 \AA\ and Fe XII 195 \AA\ lines, 
 and analyze their intensity and spectral properties. 

\section{Forward Modeling}
\label{sec_forward_model}

The intensity of a spectral line is obtained by integrating the emissivity along a line of sight (LoS)
\begin{eqnarray}
I = \int_{\rm LoS} \frac{\epsilon}{4\pi} dl~,
\label{eq_def_I}   
\end{eqnarray}
with the emissivity $\epsilon$ first calculated at each grid point in the $r-z$ plane via 
\begin{eqnarray}
\label{eq_def_eps}
\epsilon = GN^2~,
\end{eqnarray}
and then converted to the Cartesian coordinate with a spacing of $50~$km in all three directions.

The contribution function $G$ is given by 
\begin{eqnarray}
\label{eq_def_G}
G = h\nu_{ij}\cdot0.83\cdot Ab({\rm{Fe}}) f \frac{n_jA_{ji}}{N}~.
\end{eqnarray}
Here $h\nu_{ij}$ is the energy level difference,
$Ab({\rm{Fe}})$ is the abundance of Fe relative to Hydrogen,
$f$ is the ionic fraction of the Fe ion responsible for the emission line (here it refers to either Fe X or Fe XII),
$n_j$ is the fraction of ions lying in the excited state,
and $A_{ji}$ is the spontaneous transition probability.
We compute $G$ using the function \textbf{g\_of\_t} from the CHIANTI package
(\citeauthor{2015A&A...582A..56D}~\citeyear{2015A&A...582A..56D})\footnote{http://www.chiantidatabase.org/}.

The ionic fraction $f_q$ of each Fe ion is obtained by solving a coupled set of equations~\citep[e.g.,][]{2019ApJ...870...99S},
\begin{eqnarray}
 \label{eq_ionic_frac}
 \displaystyle
 \left(\frac{\partial}{\partial t}+{\mathbf v}\cdot\nabla\right) f_q
 = N \left[f_{q-1}C_{q-1}-f_q\left(C_q+R_q\right)+f_{q+1}R_{q+1}\right]~,
\end{eqnarray}
where the ionization ($C$) and recombination ($R$) rate coefficients are found from CHIANTI as well.

	By ``equilibrium ionization''(EI), we mean that the ionic fractions are found via solving the coupled set of algebraic equations
	by neglecting the left hand side of Equation~\eqref{eq_ionic_frac}.
	Note that only the terms in the square parentheses matter in this case. 
	Given that the rate coefficients are functions of the electron temperature $T$ only, 
	the derived ionic fractions ($f_q$) also depend only on $T$. 
	Note that the last term in Equation~\eqref{eq_def_G} depends essentially only on $T$, 
	one then finds that the contribution function $G$ depends essentially only on $T$.
	In the non-EI cases, on the other hand, the ionic fractions are found by solving Equation~\eqref{eq_ionic_frac} in full.
	To initiate this solution procedure, we use the EI solutions pertaining to the fluid parameters at $t=0$ as initial conditions. 
	Note that now $N$ is no longer irrelevant. 
	As a consequence, both $f_q$ and eventually $G$ will possess an $N$-dependence.

For the spectral profiles of the emission lines,   
   we evaluate, at each grid point, the monochromatic emissivity $\epsilon_{\lambda}$
   at wavelength $\lambda$ as given by~\citep[e.g.,][]{2016FrASS...3....4V}
\begin{eqnarray}
   \epsilon_\lambda = \frac{2\sqrt{2\ln2}}{\sqrt{2\pi}\lambda_w}\epsilon
      \exp\left\{
         -\frac{4\ln2}{\lambda_w^2}\left[
                                   \lambda-\lambda_0\left(1-\frac{v_{\rm LoS}}{c}\right)
                                   \right]^2
         \right\}~.
\label{eq_def_eps_lambda}
\end{eqnarray}
Here $\lambda_w=(2\sqrt{2\ln2})\lambda_0 (v_{\rm th}/c)$ is the thermal width,
   $v_{\rm th}$ is the thermal speed determined by the instantaneous temperature, 
   $\lambda_0$ is the rest wavelength, and    
   $v_{\rm LoS}$ is the instantaneous velocity projected onto an LoS.
The monochromatic intensity $I_\lambda$ is then obtained by integrating $\epsilon_\lambda$ along the LoS 
\begin{eqnarray}
   I_{\lambda} = \int_{\rm LoS} \frac{\epsilon_{\lambda}}{4\pi} dl.
\label{eq_def_I_I_lambda}   
\end{eqnarray}
We set the spectral resolution as 22 m\AA~and the spatial resolution as $1\arcsec$,
which are compatible with EIS.

\subsection{Effects of Non-equilibrium Ionization}
\begin{figure}
\begin{centering}
 \includegraphics[width=0.8\linewidth]{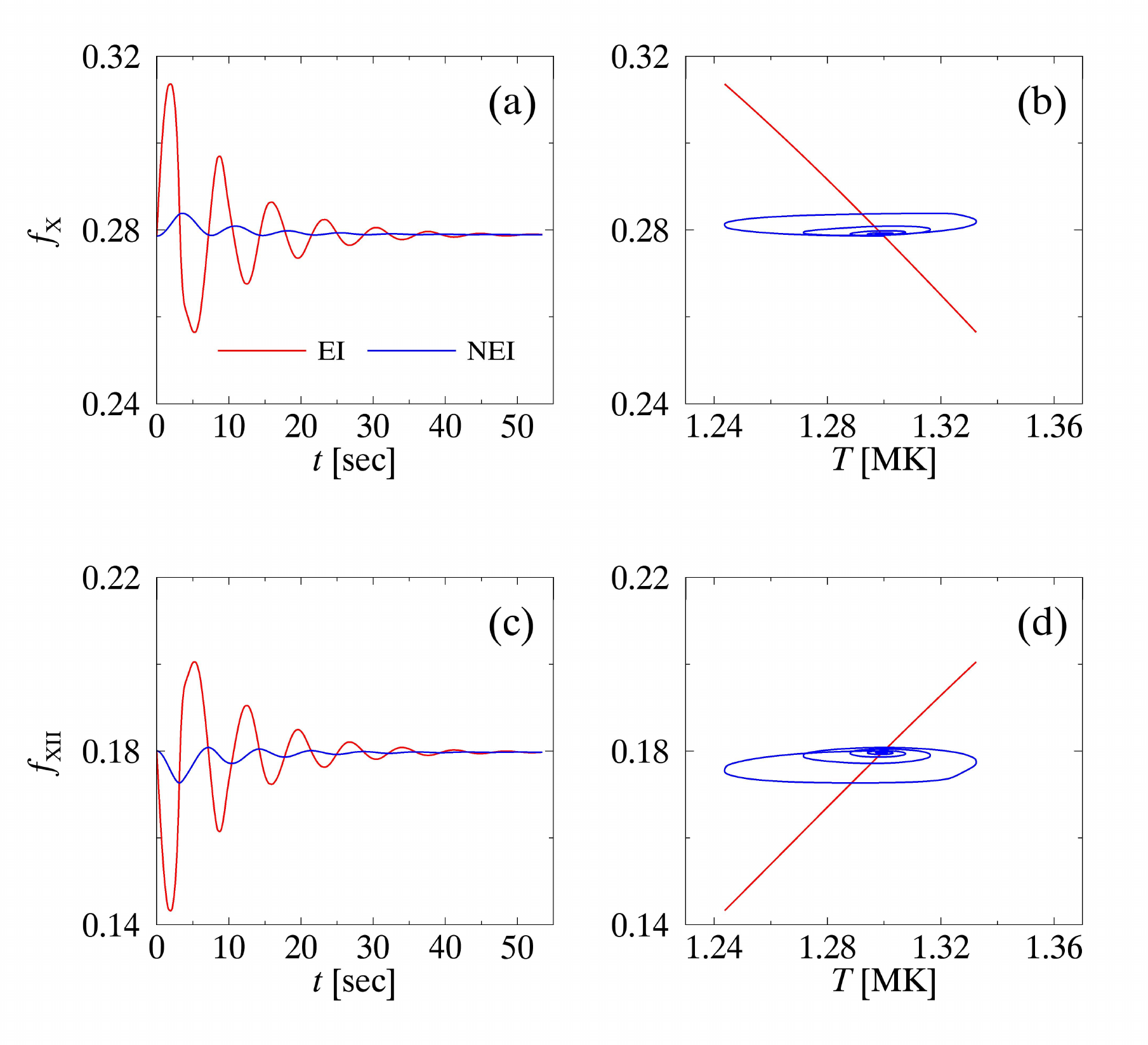}
 \caption{Ionic fractions at the loop apex of Fe X (top) and Fe XII (bottom)
 		versus time (left) and versus temperature (right).
 		The red lines are for the equilibrium ionization (EI) cases while blue lines the non-EI cases.
 		Here the base model is examined, i.e., the electron temperature at the loop axis $T_i=1.3$~MK.
 		An animation showing the trajectories is available online.
 }
 \label{f2}
\end{centering}
\end{figure}

\begin{figure}
	\begin{centering}
		\includegraphics[width=0.6\linewidth]{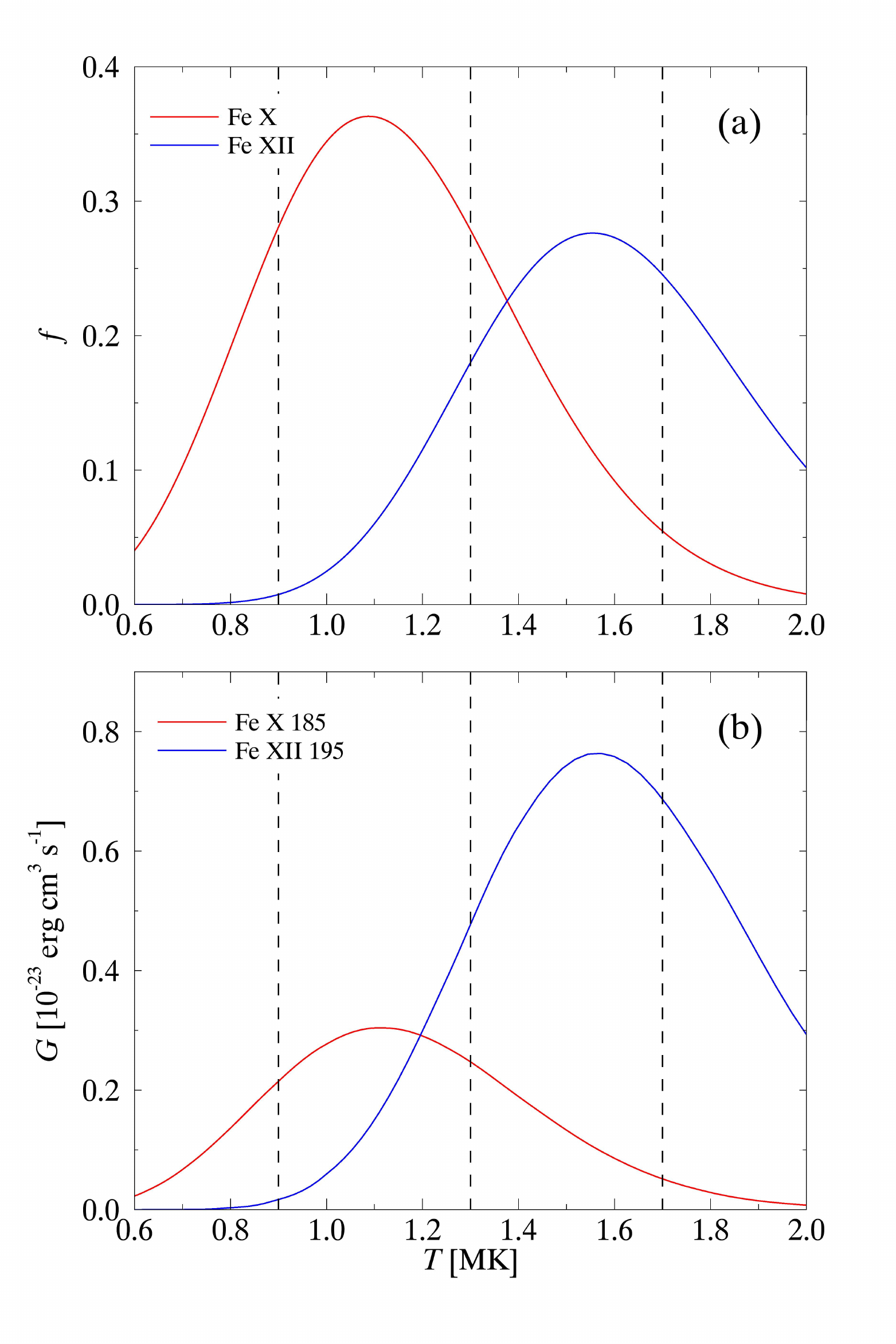}
		\caption{(a) The ionic fractions of Fe X (red) and Fe XII (blue)
			under the assumption of EI.
			(b) The contribution functions $G$
			of Fe X 185 \AA~line (red) and Fe XII 195 \AA~line (blue) in the EI case.
			Three vertical dashed lines mark the temperatures of 0.9 MK, 1.3 MK, and 1.7 MK, respectively.
		}
		\label{f3}
	\end{centering}
\end{figure}

We first show how non-equilibrium ionization (NEI) can 
	influence the ionic fractions of Fe X and Fe XII. 
Taking the base model as an example, 
Figure \ref{f2} displays the ionic fractions of Fe X (top) and Fe XII (bottom) at the loop apex ($[r,z]=[0, L_0/2]$)
	for both equilibrium ionization (EI, red lines) and non-equilibrium ionization (NEI, blue lines). 
Let us start with the temporal evolutions of the ionic fractions for (a) Fe X and (c) Fe XII.
In the cases of EI, the ionic fractions $f_{\rm{X}}$ and  $f_{\rm{XII}}$ change instantaneously with the temperature 
		(see Figure \ref{f1}a for the temperature evolution).
The opposite trends between $f_{\rm{X}}$ and $f_{\rm{XII}}$ are due to their different formation temperatures,
	which is further demonstrated in Figure \ref{f3}a.
Figure \ref{f3}a shows the ionic fractions of Fe X (red) and Fe XII (blue) 
	with respect to temperature in the EI state.
For the convenience of following discussion, 
	three vertical dashed lines indicating the temperatures of 0.9 MK, 1.3 MK, and 1.7 MK, respectively
	are overplotted in Figure \ref{f3}.  
In the base model, the background temperature at the loop apex is $1.3$~MK,
	so the ionic fraction of Fe X (Fe XII) will increase (decrease) as the temperature decreases.  
While in the cases of NEI (blue lines in Figure \ref{f2}a and \ref{f2}c),
	because of the ionization and recombination processes, 
	the variation amplitudes of the ionic fractions are dramatically decreased
	and a phase shift is seen with respect to the EI cases.
Now move to the right panels of Figure \ref{f2},
	which show the trajectories of the ionic fractions and the corresponding temperatures
	(see the animation of the trajectories).
Take $f_{\rm{X}}$ for example, in the EI case, 
	the trajectory of $f_{\rm{X}}$ starts at $[f_{\rm{X}},T]\approx[0.28,1.3~\rm{MK}]$
	and then moves back and forth along the red line.
While in the NEI case, the trajectory starts at the same location
	but moves along the blue ellipse.
Different trajectories in red and blue lines demonstrate the different responses of the ionic fractions
with respect to temperature in EI and NEI cases.
At the later stage when the sausage mode has strongly damped,
	the trajectories of $f_{\rm{X}}$ in both cases stop at their initial locations.

\begin{figure}
  \begin{centering}
  \includegraphics[width=0.8\linewidth]{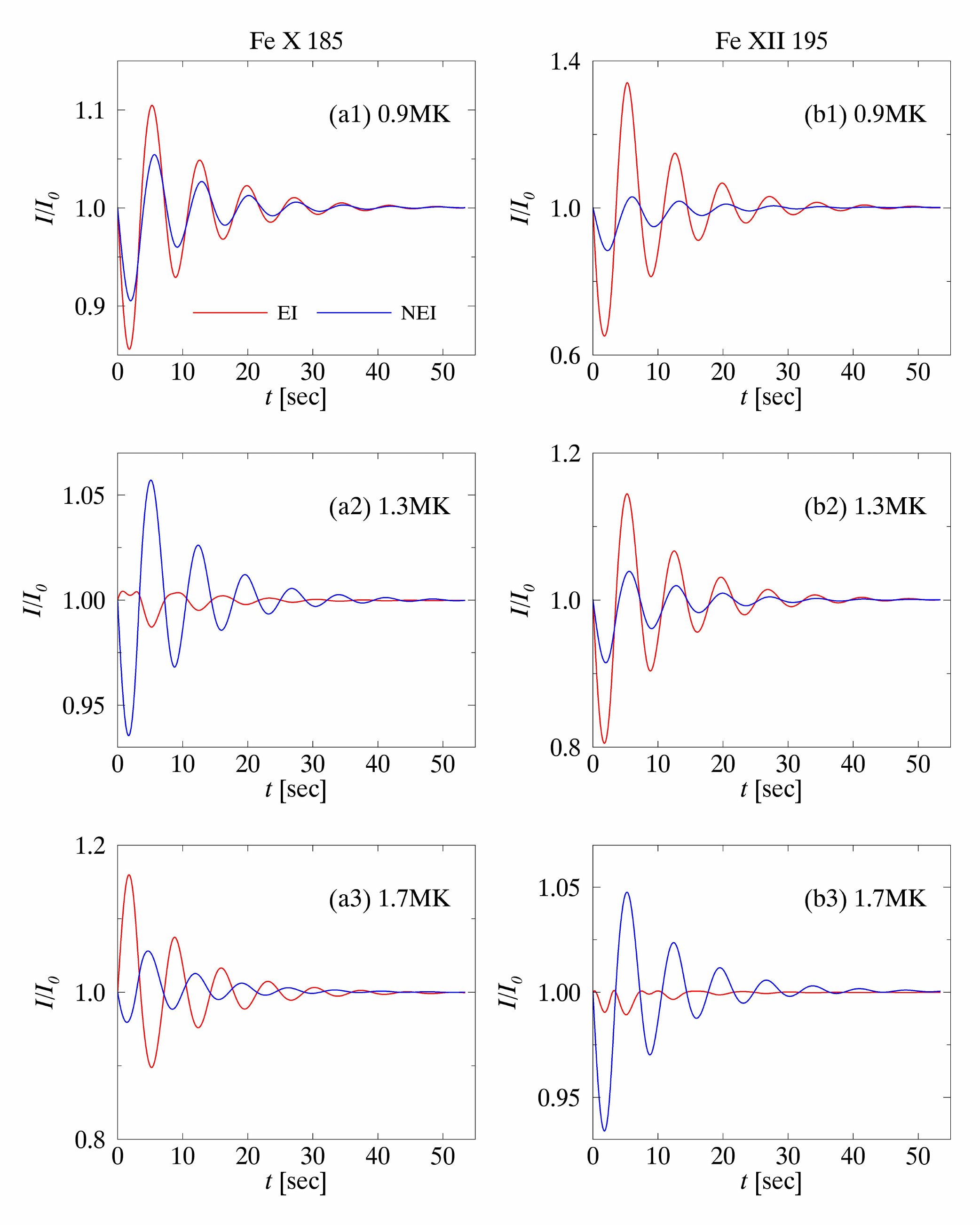}
   \caption{Temporal evolutions of the normalized intensity of Fe X 185 \AA~line (left)
   	and Fe XII 195 \AA~line (right) for simulation cases with a number of values of $T_i$,
   	the electron temperature at the loop axis.
   The LoS is perpendicular to the loop axis and passes through the loop apex.}
 \label{f4}
\end{centering}
\end{figure}

Next, for three representative cases, we show in Figure \ref{f4} the evolutions of the normalized intensity for
	the Fe X 185 \AA~line (left) and Fe XII 195 \AA~line (right),
	for both the EI (red lines) and NEI (blue lines) cases.
Here by "normalized" we mean the intensity ($I$) divided by its value at $t=0$.
The intensity is obtained by integrating the emissivity (i.e., Equation \eqref{eq_def_I})
	along the LoS that is perpendicular to the loop axis and passes through the loop apex. 
Taking the Fe X 185 \AA~line for example,
in the EI cases, quite different variations are seen 
	for different $T_i$ (see red lines in a1, a2, and a3). 
This can be explained if we analyze the emissivity variation at the loop apex.
Since the emissivity in the interior of the loop is much larger than that in the exterior of the loop,
	the intensity is largely determined by some mean column depth multiplied by the emissivity at the loop apex, i.e.,
\begin{eqnarray}
\label{equ:inten_var}
I\propto (N_0+\Delta N)^2(G_0+\Delta G)=(N_0^2+2N_0\Delta N+\Delta N^2)(G_0+\Delta G),
\end{eqnarray}
where we have omitted the column depth. Here the quantities with subscript $0$ denote the equilibrium values at the loop apex,
	while the symbols preceded by a $\Delta$ denote their variations due to the sausage mode oscillation.
We find that the intensity variations in Equation \eqref{equ:inten_var} are caused mainly by two first order terms 
$\Delta_N \equiv 2N_0\Delta NG_0$ and $\Delta_G \equiv N_0^2\Delta G$.
The two terms are due to the density variation and the contribution function variation, respectively. 
To compare the two first order terms in a simpler way, we assume that $G$ is the function only of temperature $T$.
This can be justified because the dependence of $G$ on $N$ is extremely weak.
Figure \ref{f3}b plots the contribution function $G$ versus temperature $T$ in the EI case ($N=10^9~\rm{cm^{-3}}$).
We can see that for $T_i = 0.9~\rm{MK}$, $G$ will increase if $T$ increases,
which means that the two first order terms, $\Delta_N$ and $\Delta_G$,  will change in phase,
	as $T$ and $N$ are in phase at the loop apex for FSMs.
While for $T_i = 1.3~\rm{MK}$ or $1.7$ MK, the contribution function changes oppositely 
	with respect to the density variation, 
	so the two first order terms will compete with each other.
If we calculate the ratio between $\Delta_N$ and $\Delta_G$
\begin{eqnarray}
\mathcal{R}=\frac{\Delta_N}{\Delta_G} = 2\frac{\frac{1}{N_0}\frac{dN}{dT}\Delta T}{\frac{1}{G_0}\frac{dG}{dT}\Delta T}
=2\frac{\frac{1}{N_0}\frac{dN}{dT}}{\frac{1}{G_0}\frac{dG}{dT}},
\end{eqnarray}
we find that $|\mathcal{R}|\approx1$ for $T_i$=1.3 MK, and $|\mathcal{R}|\approx0.3$ for $T_i$=1.7 MK.
This means that for $T_i = 1.3~\rm{MK}$, $\Delta_N$ and $\Delta_G$ almost cancel each other,
	which makes the intensity variation amplitude very weak.
While for $T_i = 1.7~\rm{MK}$, $\Delta_G$ is dominant over $\Delta_N$,
	so that the intensity variation is in anti-phase with respect to the density variation.
	
In the cases of NEI, the intensity variations are quite different from the EI cases.
The reason is that NEI reduces $\Delta_G$ via decreasing 
	$\partial G/\partial T$ (note that now $G$ possesses a substantial $N$-dependence as well).
So that for $T_i = 0.9~\rm{MK}$,
	 the intensity variation is weaker compared with the EI case,
and for $T_i = 1.3~\rm{MK}$ or $1.7~\rm{MK}$, $\Delta_N$ becomes dominant over $\Delta_G$.
The result is that in the NEI cases, the intensity variations are dominated by the density variation and thus show similar trend for different $T_i$.
The intensity variation of the Fe XII 195 \AA~line can be explained in a similar way.

\begin{figure}
	\begin{centering}
		\includegraphics[width=0.6\linewidth]{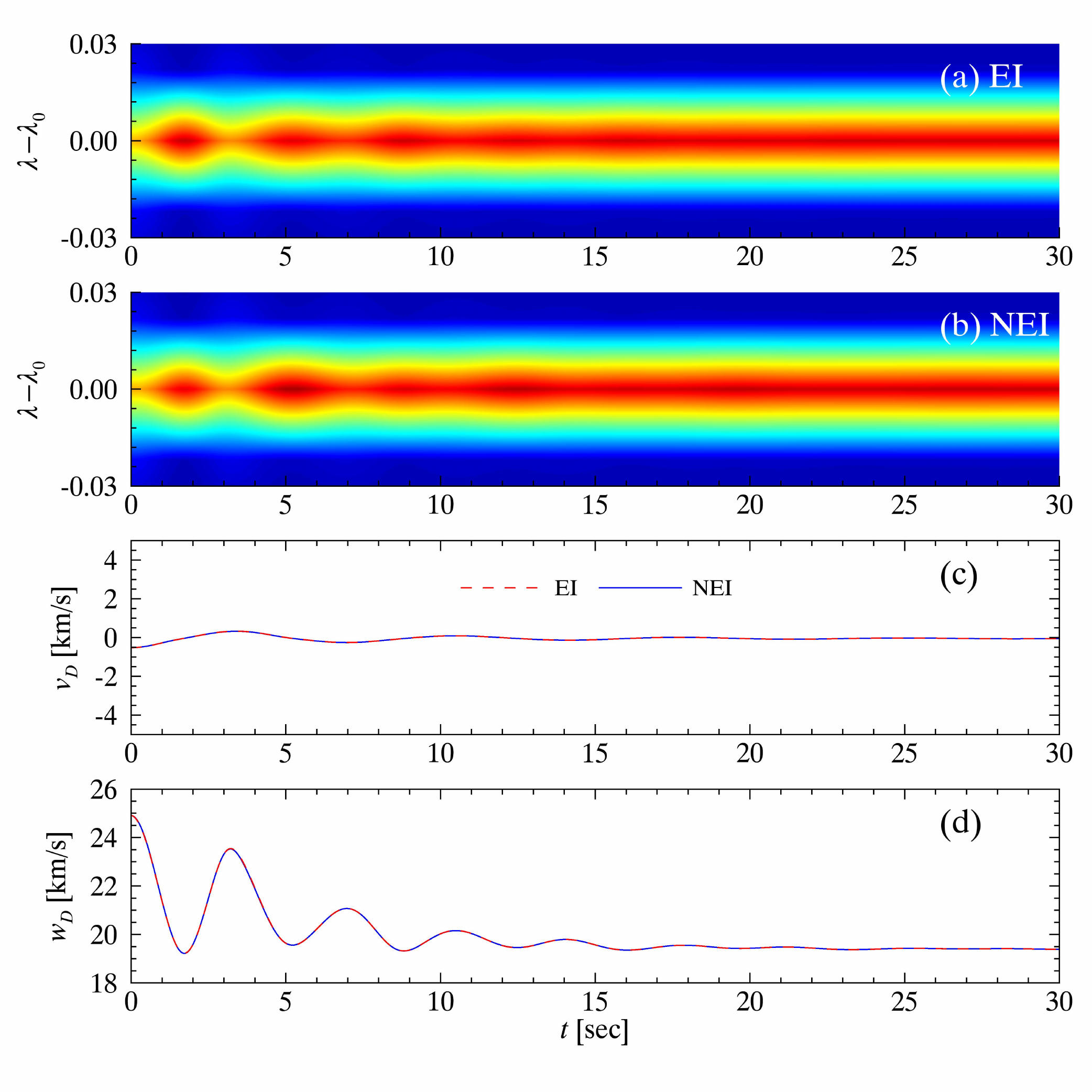}
		\caption{Spectral profiles of the Fe X 185 \AA~line for EI and NEI cases. 
			From top to bottom: spectral profiles $I_\lambda$, 
			Doppler velocity $v_D$, and Doppler width $w_D$.
		The base model is examined, i.e., the electron temperature at the loop axis $T_i=1.3$~MK.
		Here a different LoS is chosen. The LoS passes through $[r,z]=[0,L_0/4]$ and is $45^\circ$ with respect to the loop axis.}
		\label{f5}
	\end{centering}
\end{figure}

Now examine the spectral properties of the emission line,
also taking the Fe X 185 \AA~line in base model as an example.
Figure \ref{f5} shows the temporal evolutions of the monochromatic intensity $I_\lambda$
	for (a) EI and (b) NEI,
	the derived Doppler velocity $v_D$ (c), and Doppler width $w_D$ (d).
For the monochromatic intensity, the LoS passes through $[r,z]=[0,L_0/4]$ and is $45^\circ$ with respect to the loop axis.
This LoS samples the asymmetric region of the loop and is expected to have a none-zero Doppler velocity.
From Figure \ref{f5} we see that 
	even though the monochromatic intensity is slightly different in the NEI case from the EI case,
	there is no obvious difference of either the Doppler velocity or the Doppler width.
We note that the Doppler velocity and Doppler width are not that useful for deriving the wave parameters from a practical perspective.
Firstly, the Doppler velocity is weak and therefore challenging to detect.
Secondly, the period of the Doppler width is about half of the wave period,
	so it is hard to get the period or damping information from the spectral measurements.
	
It is also worth noting that the NEI cases are more realistic for observations,
especially for the FSMs possessing a short wave period.
In what follows, we only analyze the intensity variations in the NEI cases,
	and consider only the LoS that is perpendicular to the loop axis and passes through the loop apex.

\subsection{Damping of the Leaky Sausage Modes}
\begin{figure}
	\begin{centering}
		\includegraphics[width=0.6\linewidth]{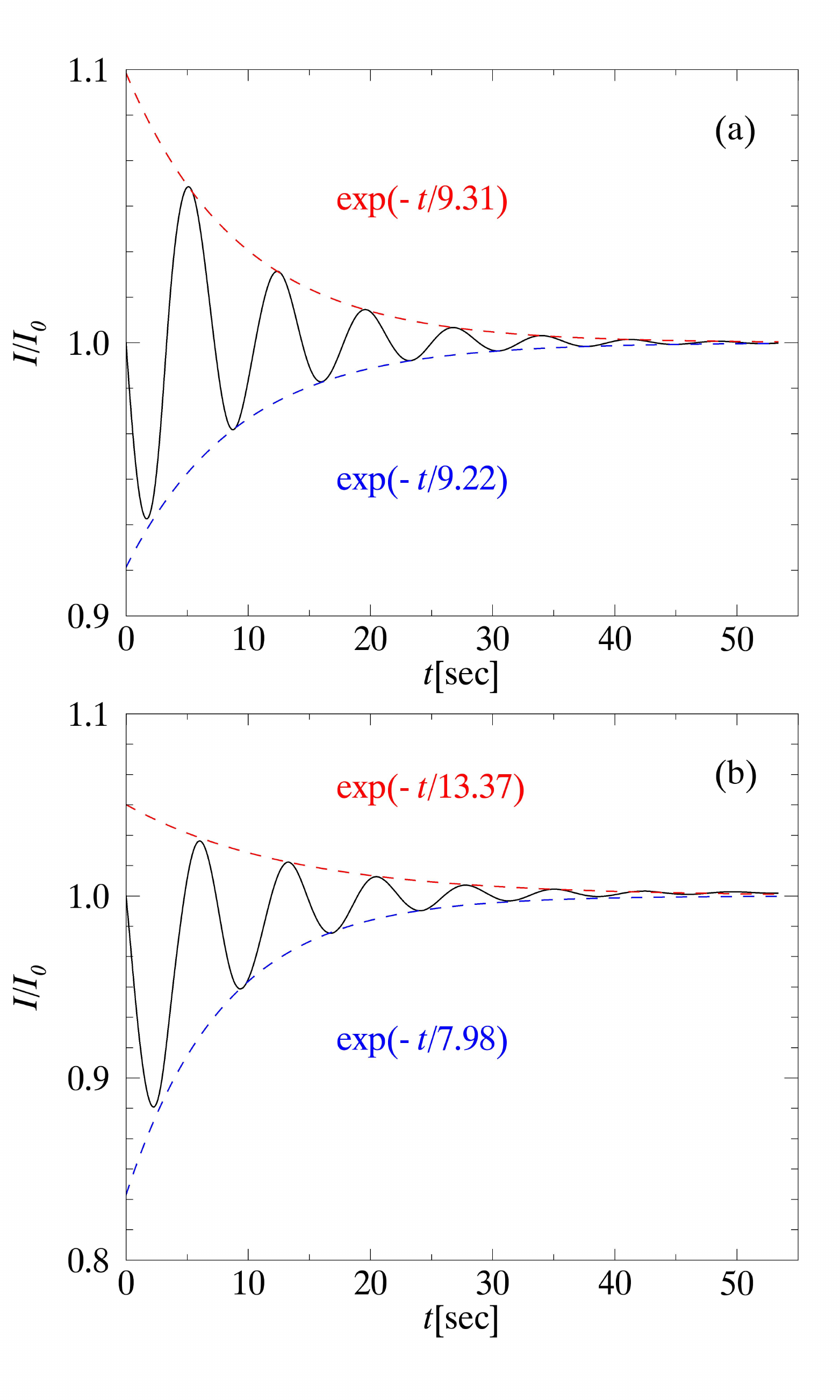}
		\caption{Exponential damping fits using the crests (red dashed lines)
			and the troughs (blue dashed lines) for the NEI intensity variations of 
		(a) Fe X 185 \AA~with $T_i=1.3~\rm{MK}$, and (b) Fe XII 195 \AA~with $T_i=0.9~\rm{MK}$.
	The LoS is perpendicular to the loop axis and passes through the loop apex.}
		\label{f6}
	\end{centering}
\end{figure}

Even though the fluid parameters of the leaky FSMs are well defined 
	exponential damping curves (see Figure \ref{f1}),
	the intensity variations can be anti-symmetric between the crests and the troughs.
To demonstrate this, in Figure \ref{f6}, 
	we replot two NEI cases from Figure \ref{f4} together with their exponential damping fits 
	using the crests and troughs separately.
In Figure \ref{f6}a, the damping times derived from the crests and troughs are very close
	and both agree well with the wave damping time.
While in Figure \ref{f6}b, the damping times from the crests and troughs 
	differ substantially and both show obvious deviations with respect to the wave damping time.

Using the same method, we derive the damping times from the intensity variations for other cases 
	and summarize the results in Table \ref{tab}.
The first row of Table \ref{tab} shows the simulation cases,
	followed by the damping time $\tau$ derived from the fluid parameter $v_r$ in the second row.
In the third row, we display the damping time from the intensity variations of the Fe X 185 \AA,
	with the values derived from the crests at the top and from the troughs at the bottom.
The last row is the same as the third row but from the Fe XII 195 \AA~line.
The numbers in the parentheses show relative errors 
	with respect to the FSMs damping time (i.e., the second row).
From Table \ref{tab}, we find that in some cases, the damping time deviates considerably from the wave damping time,
	while in other cases, the damping time is in close agreement with the wave damping time.
Furthermore, the relative errors of the damping time show systematical changes with respect to $T_i$.

\begin{figure}
	\begin{centering}
		\includegraphics[width=0.6\linewidth]{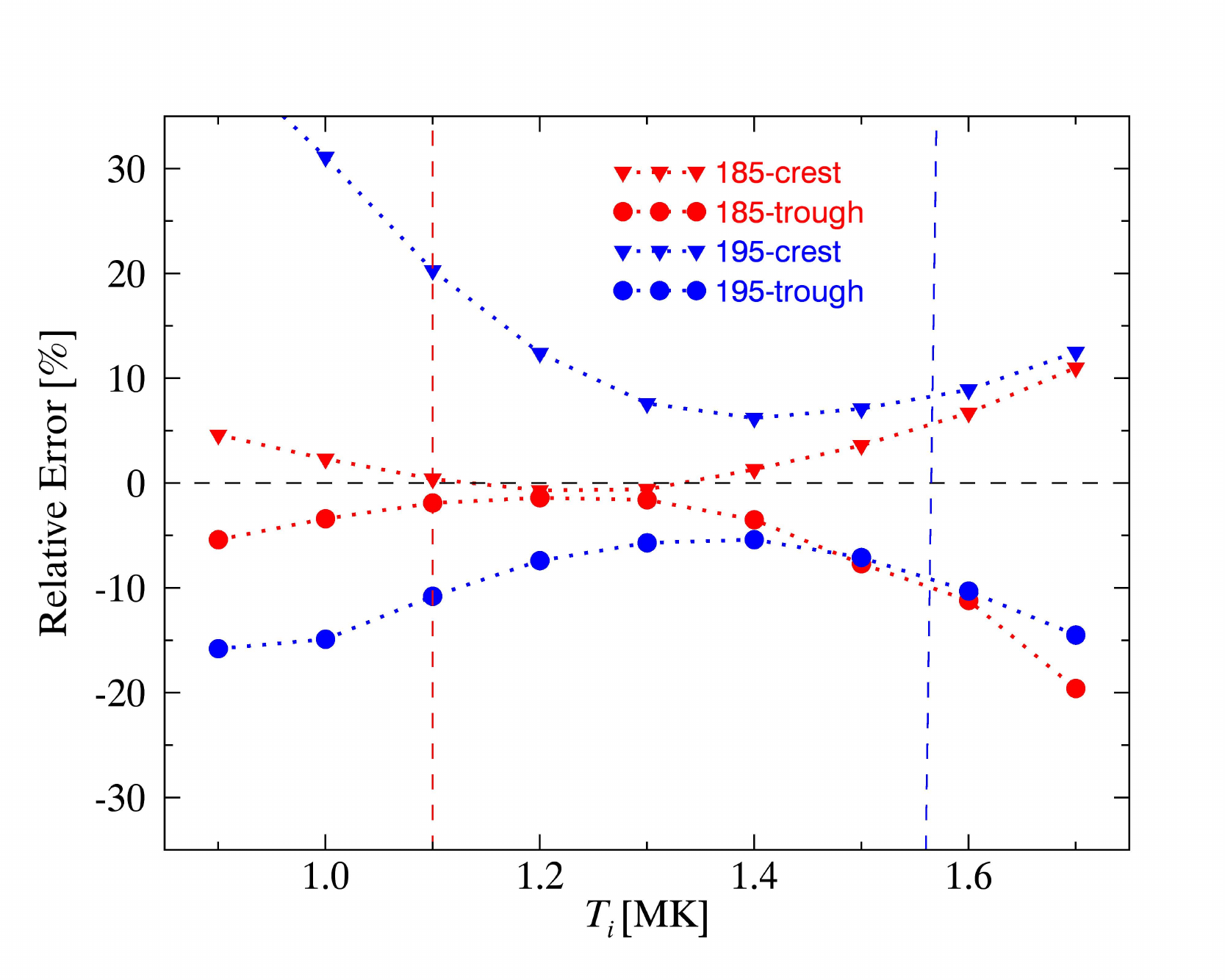}
		\caption{Relative errors of the damping time from the intensity variations with respect to the damping time of FSMs (see Table \ref{tab} for details).
		Two vertical lines mark the nominal formation temperatures of the Fe X 185 \AA~and Fe XII 195 \AA~lines.
	The LoS is perpendicular to the loop axis and passes through the loop apex.}
		\label{f7}
	\end{centering}
\end{figure}

To further demonstrate this trend, we plot in Figure \ref{f7} the relative errors of the damping times.	
Figure \ref{f7} shows that when $T_i$ is around the nominal formation temperatures of the spectral lines (vertical dashed lines),
	the relative error is small,
	while when $T_i$ deviates a lot from the nominal formation temperatures, the relative error can be large.
The reason for this trend is that when $T_i$ is around the nominal formation temperature,
	the first order term $\Delta_G$ is considerably reduced,
	so that the intensity variations are mostly caused by the density variation term $\Delta_N$,
	which makes the damping of the intensity largely follow the damping the FSMs. 
We conclude that when the loop temperature is around the nominal formation temperature of a spectral line,
	the damping time derived from the intensity can largely reflect the damping time of the leaky FSMs.
While when the loop temperature is quite different from the nominal formation temperature,
	caution need to be exercised when interpreting the damping time from the intensity variations.

\begin{deluxetable*}{cccccccccc}
	\tabletypesize{\scriptsize}
	\tablewidth{0pt} 
	\tablenum{1}
	\tablecaption{Damping times of the leaky FSMs from the fluid parameters $v_r$
		and from the intensity variations of the Fe X 185 \AA~and Fe XII 195 \AA~lines.  
		\label{tab}}
	\tablehead{
		\colhead{$T_i$ [MK]} & \colhead{0.9} & \colhead{1.0}& \colhead{1.1} & \colhead{1.2} &
		\colhead{1.3} & \colhead{1.4} & \colhead{1.5} & \colhead{1.6} & \colhead{1.7}			
	} 
	\startdata 
	MHD [sec]     & 9.48          & 9.44           & 9.42          & 9.40           & 9.37          & 9.34     & 9.32 & 9.30 & 9.28 \\
	\cline{1-10}
	Fe X 185      & 9.92(4.6\%)  & 9.66(2.3\%)   & 9.46(0.4\%)  & 9.33(-0.7\%)  & 9.31(-0.6\%) & 9.46(1.3\%) & 9.66(3.6\%)  & 9.92(6.7\%)  & 10.30(11\%)  \\
	{[sec]}       & 8.97(-5.4\%) & 9.12(-3.4\%)  & 9.24(-1.9\%) & 9.27(-1.4\%)  & 9.22(-1.6\%) & 9.01(-3.5\%)& 8.60(-7.7\%) & 8.26(-11.2\%)& 7.46(-19.6\%)   \\
	\cline{1-10}
	Fe XII 195    & 13.37(41\%)  & 12.38(31.1\%) & 11.33(20.3\%)& 10.57(12.4\%) &10.08(7.6\%)  & 9.92(6.2\%) & 9.98(7.1\%)  & 10.13(8.9\%) & 10.44(12.5\%)\\
	{[sec]}       & 7.98(-15.8\%)& 8.03(-14.9\%) & 8.40(-10.8\%)& 8.70(-7.4\%)  & 8.84(-5.7\%) & 8.84(-5.4\%)& 8.66(-7.1\%) & 8.34(-10.3\%)& 7.93(-14.5\%)  
	\enddata
	\tablecomments{The first row displays the simulation cases,
		follwed by the damping times derived from the MHD fluid parameter $v_r$ in the second row.
		The third row shows the derived damping time from crests (top) and troughs (bottom) of the Fe X 185 intensity variations.
		The fourth row is the same as the third row but for Fe XII 195.
		The nominal formation temperatures of Fe X 185 \AA~and Fe XII 195 \AA~lines are 1.1 MK and 1.57 MK, respectively.}   
\end{deluxetable*}

\section{Summary and Conclusion}
\label{sec:summary_and_conclusion}
In this work, we synthesize the EUV emissions modulated by leaky fast sausage modes (FSMs) in solar active region loops
	and explore their observational signatures using the spectrograph like EIS.
Starting with the MHD simulated data of standing leaky FSMs,
	we first compute the emissivity of the Fe X 185 \AA~and Fe XII 195 \AA~lines,
	and then integrate the emissivity along a line of sight to get the line intensity and spectral properties.
Non-equilibrium ionization (NEI) is incorporated in this procedure 
	via solving the ionization-recombination equations.

We find that while NEI can dramatically change the line intensity,
it has little influence to either the Doppler velocity or Doppler width.
We also find that the damping time derived from the intensity variations can largely reflect the wave damping time 
	if the loop temperature is around the nominal formation temperature of the spectral line,
while obvious differences are seen from the damping time of intensity 
	if the loop temperature deviates from the nominal formation temperature.
These results are helpful to understand the properties of EUV emissions by the leaky FSMs and 
	to identify the FSMs in solar active region loops.
	
	Before closing, some remarks seem in order to address the potential impacts of non-ideal effects on the FSMs. 
	This is because, being strongly compressible, FSMs are expected to be affected by such effects as electron thermal conduction, proton viscosity, 
	and the misbalance between volumetric heating and radiative cooling. 
	However, it seems that so far only the importance of electron thermal conduction and proton viscosity has been qualitatively assessed 
	relative to lateral leakage in attenuating FSMs~\citep{2007AstL...33..706K}. 
	While the conclusion therein is that lateral leakage plays a more important role, 
	it cannot be safely generalized to the loop parameters adopted in this manuscript. 
	On the other hand, the recent studies by \citet{2019arXiv190707051K} and \citet{2019arXiv190708168Z} 
	demonstrated that the heating/cooling misbalance may impact substantially the dispersive properties of slow waves. 
	Given the strong compressibility of FSMs, as is the case for slow waves, 
	it will be informative to examine how the heating/cooling misbalance affects FSMs as well. 
	It is therefore worth examining how these afore-mentioned non-ideal effects affect FSMs, and consequently how they affect the (E)UV emissions modulated by FSMs. Such a study is beyond the scope of this manuscript, though.
	
\acknowledgments
This work is supported by
    the National Natural Science Foundation of China (11761141002, 41674172, U1831112, 41604145, 41474149).
Z.H. is supported by the Young Scholar Program of Shandong University Weihai (2017WHWLJH07).
This work is also supported by the Open Research Program of the
Key Laboratory of Solar Activity of National Astronomical
Observatories of China (KLSA201908, KLSA201801).
We also acknowledge the International Space Science Institute Beijing (ISSI-BJ) for 
	supporting the international teams “MHD Seismology of the Solar Corona,” and “Pulsations in solar flares:
	matching observations and models.”  
CHIANTI is a collaborative project involving George Mason University, 
    the University of Michigan (USA) and the University of Cambridge (UK).

\bibliographystyle{apj}
\bibliography{test}




\end{document}